\pdfoutput=1
\documentclass[11pt,twoside,a4paper,cmspaper,final,collab]{cms-tdr}
\def\svnVersion{42de2ce}\def\svnDate{2021/09/22}\def\cmsCernNoTag{CERN-EP-2021-125}\def\cmsCernDate{\today}\def\cmsMessage{Published in Physical Review Letters as \href{http://dx.doi.org/10.1103/PhysRevLett.127.261804}{\doi{10.1103/PhysRevLett.127.261804}.}}
\begin{document}\cmsNoteHeader{EXO-20-015}

\newlength\cmsTabSkip\setlength\cmsTabSkip{1ex}
\newlength\cmsFigWidth
\ifthenelse{\boolean{cms@external}}{\setlength\cmsFigWidth{0.48\textwidth}}{\setlength\cmsFigWidth{0.6\textwidth}}
\ifthenelse{\boolean{cms@external}}{\providecommand{\cmsLeft}{upper\xspace}}{\providecommand{\cmsLeft}{left\xspace}}
\ifthenelse{\boolean{cms@external}}{\providecommand{\cmsRight}{lower\xspace}}{\providecommand{\cmsRight}{right\xspace}}
\providecommand{\cmsTable}[1]{\resizebox{\textwidth}{!}{#1}}
\ifthenelse{\boolean{cms@external}}{\providecommand{\CL}{C.L.\xspace}}{\providecommand{\CL}{CL\xspace}}
\ifthenelse{\boolean{cms@external}}{\providecommand{\NA}{\ensuremath{\cdots}\xspace}}{\providecommand{\NA}{\ensuremath{\text{---}}\xspace}}

\newcommand{\PS}{{\HepParticle{S}{}{}}\Xspace}
\newcommand{\ddbar}{\ensuremath{\PQd\PAQd}\xspace}
\newcommand{\htoss}{\ensuremath{\PH\to\PS\PS}\xspace}
\newcommand{\brhtoss}{\ensuremath{\mathcal{B}(\htoss)}\xspace}

\cmsNoteHeader{EXO-20-015}
\title{Search for long-lived particles decaying in the CMS endcap muon detectors in proton-proton collisions at \texorpdfstring{$\sqrt{s} = 13\TeV$}{sqrt(s) = 13 TeV}}

\date{\today}
\abstract{
A search for long-lived particles (LLPs) produced in decays of standard model (SM) Higgs bosons is presented. The data sample consists of 137\fbinv of proton-proton collisions at $\sqrt{s} = 13\TeV$, recorded at the LHC in 2016--2018. A novel technique is employed to reconstruct decays of LLPs in the endcap muon detectors. The search is sensitive to a broad range of LLP decay modes and to masses as low as a few \GeV. No excess of events above the SM background is observed. The most stringent limits to date on the branching fraction of the Higgs boson to LLPs subsequently decaying to quarks and $\PGtp\PGtm$ are found for proper decay lengths greater than 6, 20, and 40\unit{m}, for LLP masses of 7, 15, and 40\GeV, respectively.}

\hypersetup{
pdfauthor={CMS Collaboration},
pdftitle={Search for long-lived particles decaying in the CMS endcap muon detectors in proton-proton collisions at sqrt(s) = 13 TeV},
pdfsubject={CMS},
pdfkeywords={CMS, long-lived particles, muon detector}
}

\maketitle

Many extensions of the standard model (SM) predict
the existence of neutral, weakly-coupled particles that have a
long lifetime. These long-lived particles (LLPs) naturally arise
in models of split supersymmetry (SUSY)~\cite{Giudice:2004tc,Hewett:2004nw,ArkaniHamed:2004yi,Gambino:2005eh,Arvanitaki:2012ps,ArkaniHamed:2012gw},
SUSY with weak $R$-parity violation~\cite{Fayet:1974pd,Farrar:1978xj,Weinberg:1981wj,Barbier:2004ez},
SUSY with gauge-mediated supersymmetry breaking~\cite{GIUDICE1999419,Meade:2008wd,Buican:2008ws},
stealth SUSY~\cite{Fan:2011yu,Fan:2012jf},
hidden valley scenarios~\cite{Strassler:2006im,Strassler:2006ri,Han:2007ae},
baryogenesis triggered by weakly interacting massive particles~\cite{Cui:2011ab,Cui:2012jh,Cui:2014twa},
inelastic dark matter~\cite{PhysRevD.64.043502}, and twin Higgs models~\cite{Chacko:2005pe,Curtin:2015fna,Cheng:2015buv}.

\begin{figure}[htb]
   \centering
   \includegraphics[width=0.49\textwidth]{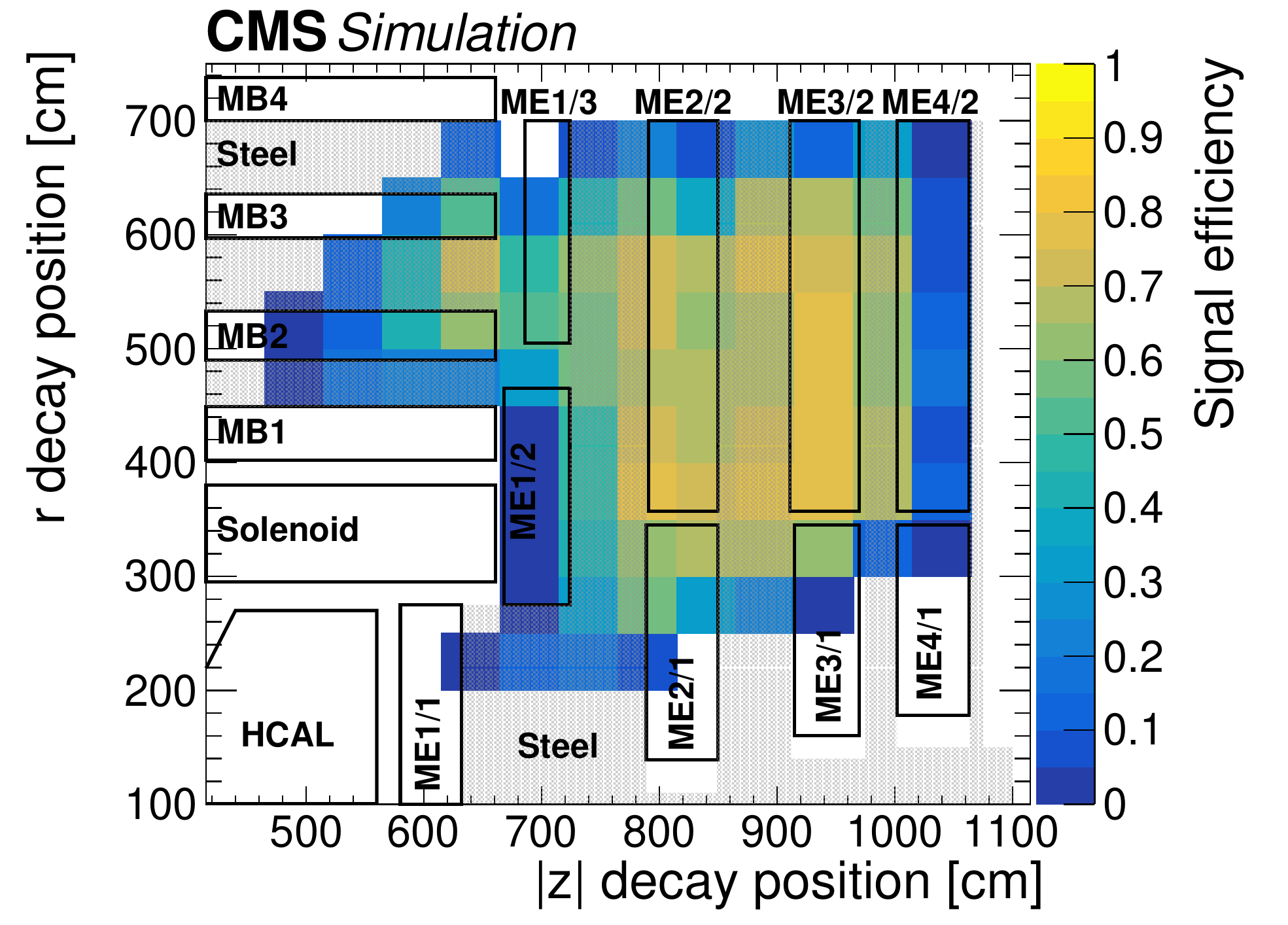}
   \caption{
   \label{fig:signal_eff_map}
  The signal efficiency of the combined cluster reconstruction, veto, and identification selections
as a function of the simulated $r$ and $z$ decay positions of \PS decaying to $\bbbar$, for a mass of
15\GeV and a uniformly distributed mixture of events with $c\tau$ between 1--10\unit{m}. The
barrel and endcap muon stations are drawn as black boxes and labeled by their station names, showing
the geometry of the muon detectors. Regions occupied by the steel return yoke are shaded in gray.
   }

\end{figure}

In this Letter, we describe the first search at the LHC that uses a muon detector
as a sampling calorimeter to identify showers produced by decays of LLPs.
The CMS endcap muon detectors (EMD) are composed of detector planes interleaved with the steel layers
of the magnet flux-return yoke. A schematic diagram of the EMD geometry
is shown in Fig.~\ref{fig:signal_eff_map}.
Decays of LLPs in the EMD induce hadronic and
electromagnetic showers, giving rise to a high hit
multiplicity in localized detector regions.
The hadron calorimeter, solenoid magnet, and steel flux-return yoke together provide 20--27 nuclear interaction
lengths of shielding, which is sufficiently large to suppress particle showers that are not
fully contained (punch-through) to negligible levels.

This search has sensitivity to singly or multiply produced LLPs decaying to final states including hadrons, taus, electrons, or photons.
The LLPs decaying to muons will very rarely produce a particle shower and will remain undetected by this search.
We focus on a benchmark simplified model motivated by the twin
Higgs scenario~\cite{Craig:2015pha,Strassler:2006im,Strassler:2006ri,Han:2007ae,Strassler:2008fv,Juknevich:2009ji}
where the SM Higgs boson ($\PH$) decays to a pair of neutral long-lived scalars (\PS),
each of which decays in turn to a pair of bottom quarks ($\PQb\PAQb$), $\PGt$ leptons ($\PGtp\PGtm$),
or down quarks (\ddbar). Further details of the simplified model can be found in Ref.~\cite{Sirunyan:2020cao}.
The most stringent previous limit for mean proper decay lengths $c\tau < 0.3\unit{m}$ is based on a search for displaced jets
in the CMS tracker~\cite{Sirunyan:2020cao}. For $c\tau > 0.3\unit{m}$, displaced vertices
in the ATLAS muon spectrometer~\cite{Aaboud:2018aqj, Aad:2019xav} set the most stringent previous limit.

There are two inherent advantages of using the CMS EMD as a sampling calorimeter 
to detect LLPs over past searches that employ displaced vertices: (i) the absorber 
material in front of the EMD acts as a shield to maintain a sufficiently low 
level of background with the detection of a single LLP decay, a level that 
current displaced vertex searches can only achieve by requiring the detection of
two LLP decays; (ii) the calorimetric nature of the particle shower, sensitive 
to the LLP energy rather than its mass, renders this search equally sensitive 
to all LLP masses considered, while vertex reconstruction efficiency
tends to decrease with the LLP mass due to increasingly smaller opening angles.

Because of these advantages, the signal acceptance and sensitivity are improved relative to
the previous best results~\cite{Aaboud:2018aqj, Aad:2019xav} by more than a factor of 6 (2)
for an LLP mass of 7\GeV ($\ge$15)\GeV and $c\tau > 100\unit{m}$. The improved
signal acceptance that results from not requiring a second detected LLP decay is particularly
powerful at very large $c\tau$, making the present search the first to be sensitive
to LLP decays with $c\tau$ up to $1000\unit{m}$ and mass between 40 and 55\GeV
for the Higgs boson to LLP decay branching fractions below $20\%$.
Tabulated results and instructions to reproduce the signal efficiency are 
provided in HEPData~\cite{hepdata}.

The search is based on proton-proton ($\Pp\Pp$) collision data at 13\TeV collected
during 2016--2018 at the CERN Large Hadron Collider (LHC), corresponding to an integrated luminosity of 137\fbinv.
The central feature of the CMS experiment is a superconducting solenoid of 6\unit{m} internal
diameter providing a magnetic field of 3.8\unit{T}.
Located within the solenoid volume are a silicon pixel and strip tracker, a lead tungstate crystal
electromagnetic calorimeter, and a brass and scintillator hadron calorimeter, 
each composed of a barrel and two endcap sections.
Muons are detected in detectors embedded in the steel flux-return yoke outside the solenoid
using three technologies: drift tubes (DTs) in the barrel, cathode strip chambers (CSCs) in the endcaps,
and resistive-plate chambers (RPCs) in the barrel and endcaps.
Further details of the CMS detector and the coordinate system definition 
can be found in Ref.~\cite{Chatrchyan:2008aa}.

The CSC detector, which covers a region of pseudorapidity between $\abs{\eta} = 0.9$ and $2.4$ and
plays a critical role for the search described in this Letter,
is shown schematically in Fig~\ref{fig:signal_eff_map}. 
It is composed of four ``stations'' in each endcap,
labeled ME1 to ME4, which are located approximately
7, 8, 9.5, and 10.5\unit{m} away from the interaction point along the beamline axis ($z$)
on both ends of the detector, and are sandwiched between steel absorbers.
Each chamber is composed of six thin layers containing
cathode strips along the radial direction and anode wires perpendicular to the strips.
Charged particles traversing the chambers ionize the gas molecules. The resulting electrons are accelerated
towards the anode wires producing an avalanche, while the positive ions travel to the opposite end and induce
signals in the cathode strips. By combining the information from signals on the anode wires and the cathode
strips of each layer, we can determine the space-time coordinates of each such ``hit''
with a resolution of 400--500\mum and 5\unit{ns}~\cite{Chatrchyan:2009af}.

The CMS event reconstruction is based on a particle-flow (PF) algorithm~\cite{Sirunyan:2017ulk},
which combines information from the tracker, calorimeters, and muon detectors to
identify charged and neutral hadrons, photons, electrons, and muons, known collectively as PF candidates,
which are clustered into jets using the anti-\kt algorithm with a distance
parameter of 0.4~\cite{Cacciari:2005hq,Cacciari:2008gp,Cacciari:2011ma}.
The transverse component of the negative vectorial sum of the momenta of all PF candidates
is the missing transverse momentum \ptvecmiss, and its magnitude is indicated as \ptmiss.
The candidate vertex with the largest value of summed physics-object $\pt^2$ is taken to be
the primary $\Pp\Pp$ interaction vertex. The physics objects are the jets, clustered using
the jet finding algorithm~\cite{Cacciari:2008gp,Cacciari:2011ma} with the tracks assigned to
candidate vertices as inputs, and the associated missing transverse momentum, taken as
the negative vector sum of the \pt of those jets.
The average neutral energy density from overlapping $\Pp\Pp$
interactions (pileup) is estimated and subtracted
from the reconstructed jet energies~\cite{Cacciari:2007fd}.

The simulated $\htoss$ signal samples are generated using
\POWHEG2.0~\cite{Nason:2004rx,Frixione:2007vw,Alioli:2010xd,Re:2010bp},
and include gluon fusion, vector boson fusion, $\PW\PH$, $\PZ\PH$, and $\ttbar\PH$ production modes.
The Higgs boson mass is set to 125\GeV, while the \PS mass ($m_\PS$) is
set to 7, 15, 40, or 55\GeV. The $c\tau$ is varied between $1\unit{mm}$ and $100\unit{m}$.
Parton showering, hadronization, and the underlying event are modeled
by \PYTHIA 8.205 and 8.230~\cite{Sjostrand:2014zea}
with parameters set by the CUETP8M1~\cite{Khachatryan:2015pea} and CP5 tunes~\cite{Sirunyan:2019dfx} used
for samples simulating the 2016 and 2017/18 datasets respectively.
The NNPDF 3.0~\cite{Ball:2014uwa} and 3.1~\cite{Ball:2017nwa} parton distribution functions
are used in the generation of all simulated samples.
The \GEANTfour~\cite{geant4} package is used to model the response of the CMS detector,
and simulated minimum-bias events are mixed with the hard interactions
in simulated events to match the observed pileup distribution in data.

An LLP that decays after it has traversed the calorimeter systems can produce
large $\ptmiss$ because its momentum will remain undetected, as
$\ptmiss$ is calculated solely from the tracker and calorimeter information.
We exploit this feature by triggering
on events with $\ptmiss > 120\GeV$~\cite{Khachatryan:2016bia},
and subsequently requiring offline $\ptmiss > 200\GeV$.
We require at least one jet with $\pt > 50\GeV$ and
pseudorapidity $\abs{\eta} < 2.4$, because
signal events passing the \ptmiss requirement are always produced together
with a jet from initial-state radiation.
To suppress backgrounds from \PW boson and top quark production, events
containing an isolated electron (muon) passing loose identification
criteria~\cite{Sirunyan:2020ycc, Sirunyan:2018fpa, Sirunyan:2019yvv}
with $\pt > 35$ (25)\GeV and $\abs{\eta} < 2.5$ (2.4) are vetoed.

The CSC hits are clustered in $\eta$ and the
azimuthal angle $\phi$ (in radians) using the \textsc{dbscan} algorithm~\cite{dbscan},
which groups hits by high-density regions.
A minimum of 50 hits and a ``distance parameter'' of 0.2 is used.
The geometric acceptance for at least one signal LLP of mass 7--55\GeV
decaying in the fiducial region of the CSC detectors
ranges from 4--27\% for $c\tau$ between 1--10\unit{m}, decreasing to
0.5--5\% for $c\tau$ of 100\unit{m}.
The efficiency for the showers to be reconstructed by the \textsc{dbscan} algorithm
is approximately 80\% for $\bbbar$ and $\ddbar$ decays and 65\% for $\PGtp\PGtm$ decays.
The accuracy of the simulation prediction for the cluster reconstruction efficiency
relies on its ability to model correctly the response of the muon detectors
in an environment with multiple particles, each producing a large number
of secondary shower particles.
This aspect is validated by measuring clusters produced in
$\PZ\to\MM$ data events where one of the muons undergoes bremsstrahlung
in the EMD and the associated photon produces an electromagnetic shower,
and the associated systematic uncertainty is taken into account in the
overall uncertainty on the reconstruction efficiency.

The main SM backgrounds include punch-through jets,
muons that undergo bremsstrahlung, and decays of SM LLPs, such as the neutral kaon \PKzL.
To suppress background from punch-through jets or muon bremsstrahlung,
we reject clusters that have a jet or muon within
a $\Delta R = \sqrt{\smash[b]{(\Delta\eta)^2+(\Delta\phi)^2}} < 0.4$ cone
and with jet $\pt > 10\GeV$ or muon $\pt > 20\GeV$.
We veto clusters that have any hits in the
two innermost rings of the ME1 station (ME1/1 and ME1/2), which have
the least absorber material in front, or
match any hit (with $\Delta R(\text{cluster, hit})<0.4$) in the
RPCs located immediately next to ME1/2.
In the region where the barrel and endcap muon detectors overlap ($0.8 < \abs{\eta} < 1.2$),
we veto any cluster matched to any track segment
reconstructed in the innermost station of the DT detectors (MB1),
or any hit in the RPCs situated in front of and behind MB1 matched to within $\Delta R(\text{cluster, segment/hit})<0.4$.
Finally, we reject clusters with $\abs{\eta} > 2.0$ to suppress the muon bremsstrahlung background
that evaded the muon veto because of the high muon reconstruction and identification
inefficiencies at larger $\abs{\eta}$.

The inefficiencies for the jet, ME1, MB1, and RPC hit vetoes
are predominantly caused by the presence of pileup particles
and random noise, and are measured by randomly
sampling the ($\eta$,$\phi$) locations of clusters from the signal
distribution and evaluating whether a jet or ME1/MB1/RPC hit has been observed
within a $\Delta R<0.4$ cone about the cluster's location,
using $\PZ\to\MM$ data events with the two muons
from the \PZ decay removed. The loss of efficiency due to the muon veto is also affected by
muon segments produced by particles resulting from the LLP decay itself. This
contribution is further validated using a control sample of
clusters matched to trackless jets made to resemble the signal LLP decay
by requiring the neutral energy fraction to be larger than 95\%.
A 10\% correction is applied to the signal efficiency
to account for the simulation's mismodeling of the vetoes.
To suppress noncollision backgrounds, we apply filters that remove events containing
beam-halo muons or calorimeter noise~\cite{Chatrchyan:2011tn}.
To suppress background from cosmic ray muon showers,
which produce hits in multiple regions of the CMS detector,
we reject any event in which more
than a quarter of DT and CSC $\phi$-rings contain 50 or more hits.
The efficiency for signal events to pass the ME1/1 and ME1/2 vetoes
depends on the LLP decay location, and its average value is between 30--60\% for the signal models considered. The efficiency
for the remaining signal to pass all other vetoes is $\sim$80\%.

After the veto requirements are applied, the dominant background source consists
of decays of SM LLPs, which are predominantly produced
by pileup interactions and are independent of the primary interaction that yielded the large \ptmiss.
These pileup interactions can occur concurrently with the primary interaction (in-time pileup)
or in adjacent bunch crossings (out-of-time or OOT pileup).
Clusters produced by OOT pileup are rejected by requiring the cluster time ($t_\text{cluster}$),
defined as the average time of the hits in the cluster relative to the LHC clock,
to be consistent with an in-time interaction ($-5.0$ $< t_\text{cluster} < 12.5$\unit{ns}).
A larger time window is used at positive values to capture signal clusters with longer delays
from slower moving LLPs. The time window requirement suppresses the background
by a factor of 5. An OOT validation region (VR) is defined
by selecting events containing clusters with $t_\text{cluster} \le -5\unit{ns}$,
and is used to validate the background prediction method.
To reject clusters composed of hits from multiple
bunch crossings, we require that the root mean square
spread of a cluster's hit times is less than 20\unit{ns}.

There are several features that distinguish between signal and background clusters.
Clusters from all background processes occur more often at larger values of $\abs{\eta}$,
as the effectiveness of the jet and muon vetoes decrease because of increasing 
reconstruction inefficiencies. Signal clusters often occupy more than one CSC station and
occur more frequently in stations further away from the primary interaction point.
A cluster identification algorithm was devised that makes successively more restrictive
$\abs{\eta}$ requirements as the number of CSC stations containing hits 
and the distance between the station and the primary interaction point decrease. 
The cluster identification algorithm has $\sim$80\% efficiency for clusters 
originating from \PS decays in the simulation, and suppresses the background by a factor of 3.

The events that pass the cluster identification criteria are used to define the search region,
and those that fail are used as an additional in-time VR.
The signal efficiency of the combined cluster reconstruction, veto, and identification selections
is shown as a function of the simulated $r$ and $z$ decay positions
of the particle \PS in Fig.~\ref{fig:signal_eff_map}. The combined efficiency averaged
over the full region of the CSC detector is 15--30\% for the signal models considered.

\begin{figure}[htb]

   \centering
   \includegraphics[width=0.49\textwidth]{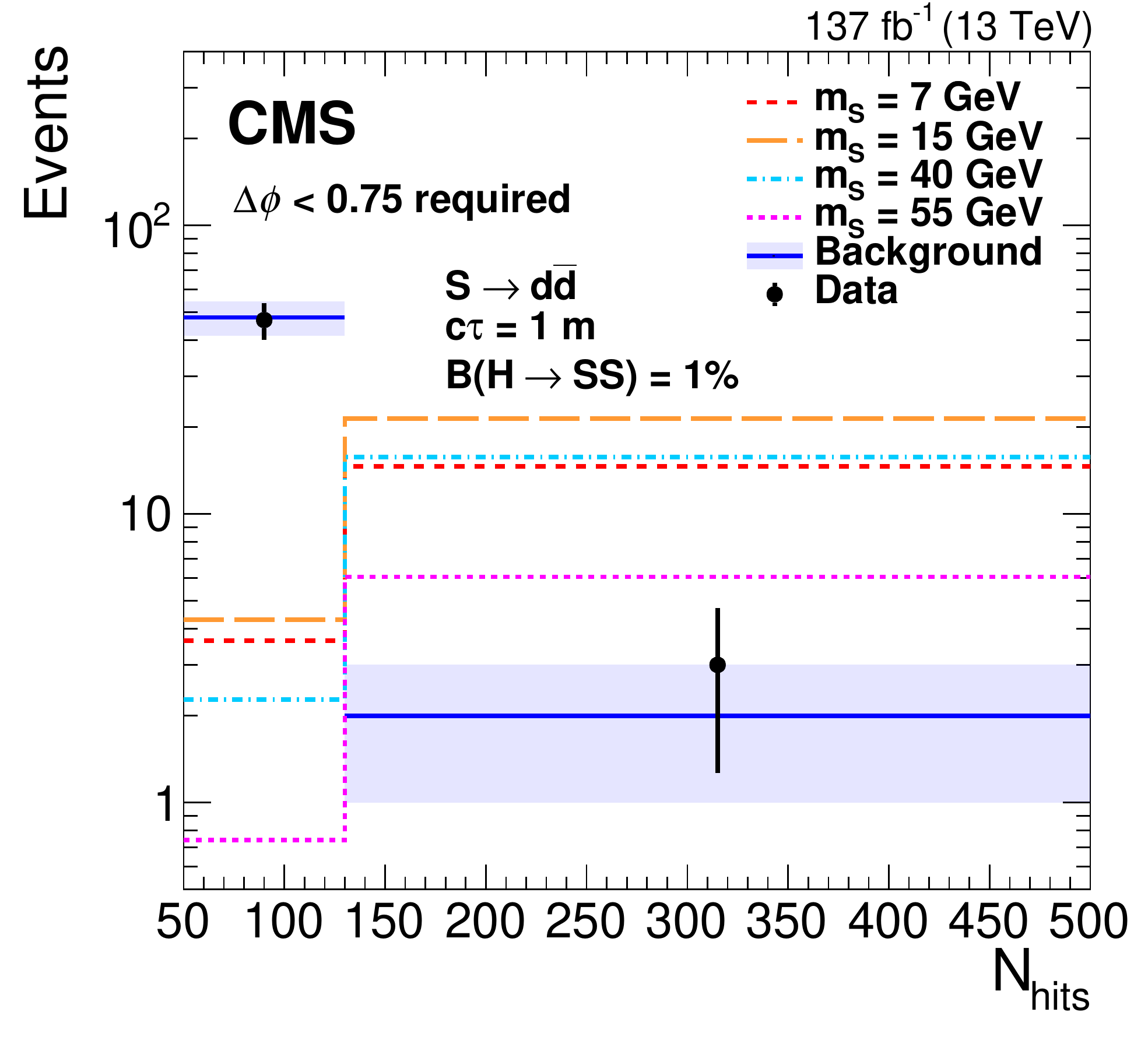}
   \includegraphics[width=0.49\textwidth]{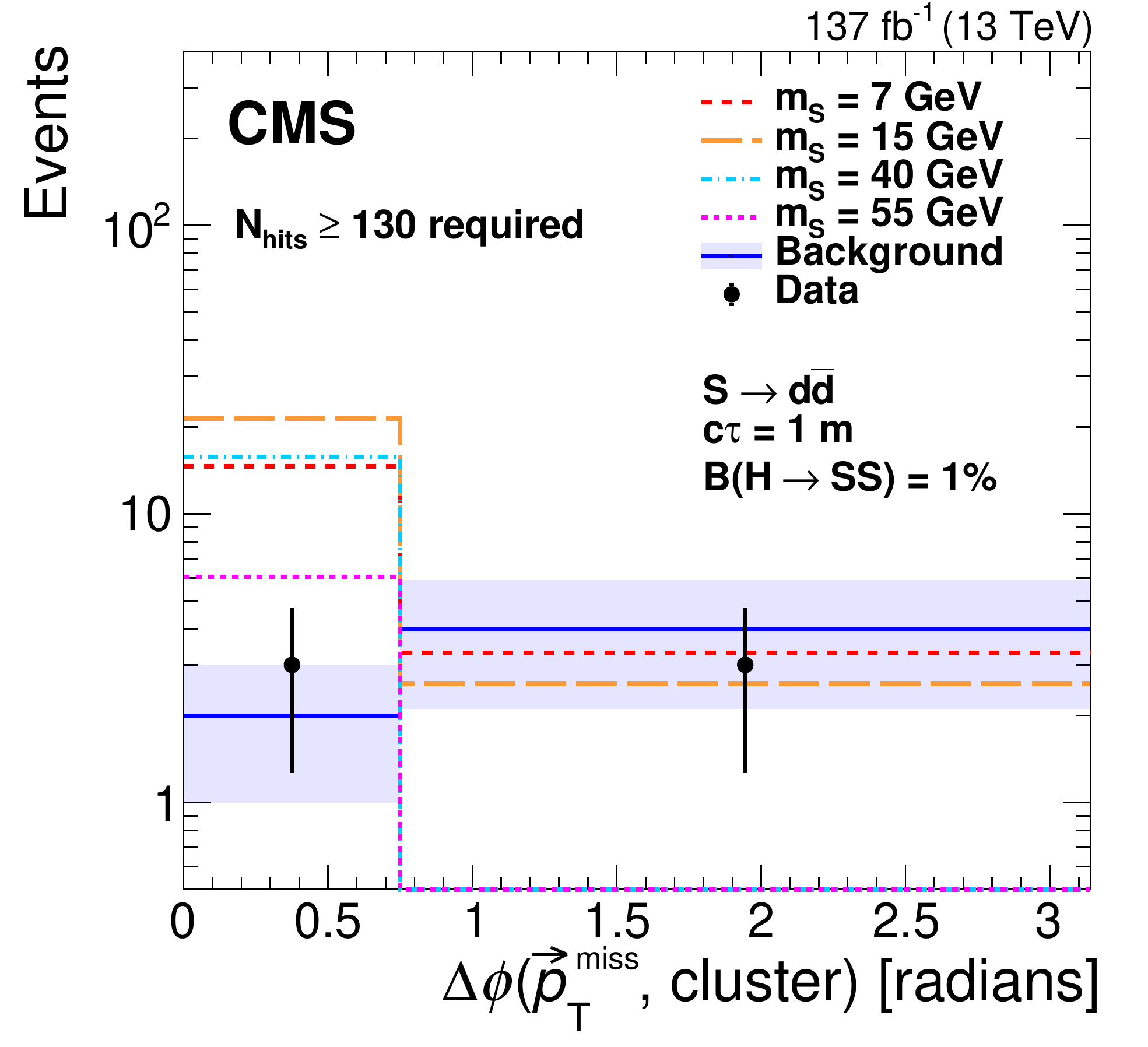}

   \caption{\label{fig:eff_signal}
   Distributions of $N_\text{hits}$ (\cmsLeft)
   and $\Delta\phi_{\mathrm{c}}$ (\cmsRight) in the search region.
   The background predicted by the fit is shown in blue
   with the shaded region showing the fitted uncertainty.
   The expected signal with $\brhtoss = 1\%$, $\PS\to\ddbar$, and $c\tau=1\unit{m}$
   is shown for $m_\PS$ of 7, 15, 40, and 55\GeV in various colors and dotted
   lines. The $N_\text{hits}$ distribution includes only events in bins C and D,
   while the $\Delta\phi_{\mathrm{c}}$ includes only events in bins A and D.
   The last bin in the $N_\text{hits}$ distributions includes overflow events.
   }
\end{figure}

The number of hits in the cluster ($N_\text{hits}$) and
 the azimuthal angle between the cluster location and the \ptvecmiss ($\Delta\phi_{\mathrm{c}}$)
are used to make the final discrimination between signal and background.
The distribution of $N_\text{hits}$ remains high at large $N_\text{hits}$
values for signal events, but for background events the distribution of
$N_\text{hits}$ decreases sharply with increasing $N_\text{hits}$ values.
For signal, $\Delta\phi_{\mathrm{c}}$ peaks near zero either because
the \ptvecmiss results from the same \PS decay that produced the cluster
or the large $\ptmiss$ requirement tends to select highly boosted Higgs bosons
for which the \PS and \PH momentum vectors are spatially close to each other.
For the backgrounds, $\Delta\phi_{\mathrm{c}}$ is
independent of $N_\text{hits}$, enabling the use of
the matrix (ABCD) method to predict the background yield in the signal-enriched
bin D as $N_\mathrm{D} = (N_\mathrm{A}N_\mathrm{C})/N_\mathrm{B}$,
where $N_\mathrm{X}$ is the background event yield in each bin X.
Bin A includes events with $\Delta\phi_{\mathrm{c}} \ge 0.75$ and $N_\text{hits} > 130$;
bin B includes events with $\Delta\phi_{\mathrm{c}} \ge 0.75$ and $N_\text{hits} \le 130$;
bin C includes events with $\Delta\phi_{\mathrm{c}} < 0.75$ and $N_\text{hits} \le 130$;
and bin D includes events with $\Delta\phi_{\mathrm{c}} < 0.75$ and $N_\text{hits} > 130$.
The distributions of $N_\text{hits}$ in bins C and D, and
$\Delta\phi_{\mathrm{c}}$ in bins A and D
are shown in Fig.~\ref{fig:eff_signal}, for the data and for the signal
assuming $\PS\to\ddbar$ decays with various \PS masses.

To account for a potential signal contribution to bins A, B, and C, a
binned maximum likelihood fit is performed simultaneously in the four bins,
with a common signal strength parameter scaling the signal yields in each bin.
The background component of the fit is constrained to obey the ABCD relationship.
Systematic uncertainties that affect the signal yield are
missing higher order QCD corrections (21\%), cluster reconstruction and identification
efficiency (6\%), veto efficiencies (4\%), jet energy scale (4\%)~\cite{Khachatryan:2016kdb},
simulation sample statistical uncertainties (3--5\%),
and luminosity (1.6\%)~\cite{REFLUMI16,REFLUMI17,REFLUMI18}. These
systematic uncertainties and the statistical uncertainty of
the simulated signal samples are treated as nuisance parameters
in the fit.

The background estimation procedure is validated using events in the OOT
and in-time VRs, predicting $1.3 \pm 0.6$ and $1.4 \pm 0.6$ events respectively. In both
VRs, 2 events are observed. In the signal-depleted A, B, and C
bins of the search region, we observe 3, 96, and 47 events
in the data, respectively. Using the fit procedure described above and
assuming no signal contribution, we predict $2.0 \pm 1.0$ background events
in the signal-enriched region D, and observe 3 events. The uncertainty
in the background prediction is dominated by the statistical uncertainty
in the event yields of the signal-depleted A, B, and C bins.
No excess of events above the SM background is observed.

\begin{figure}[htb]

  \centering
  \includegraphics[width=0.49\textwidth]{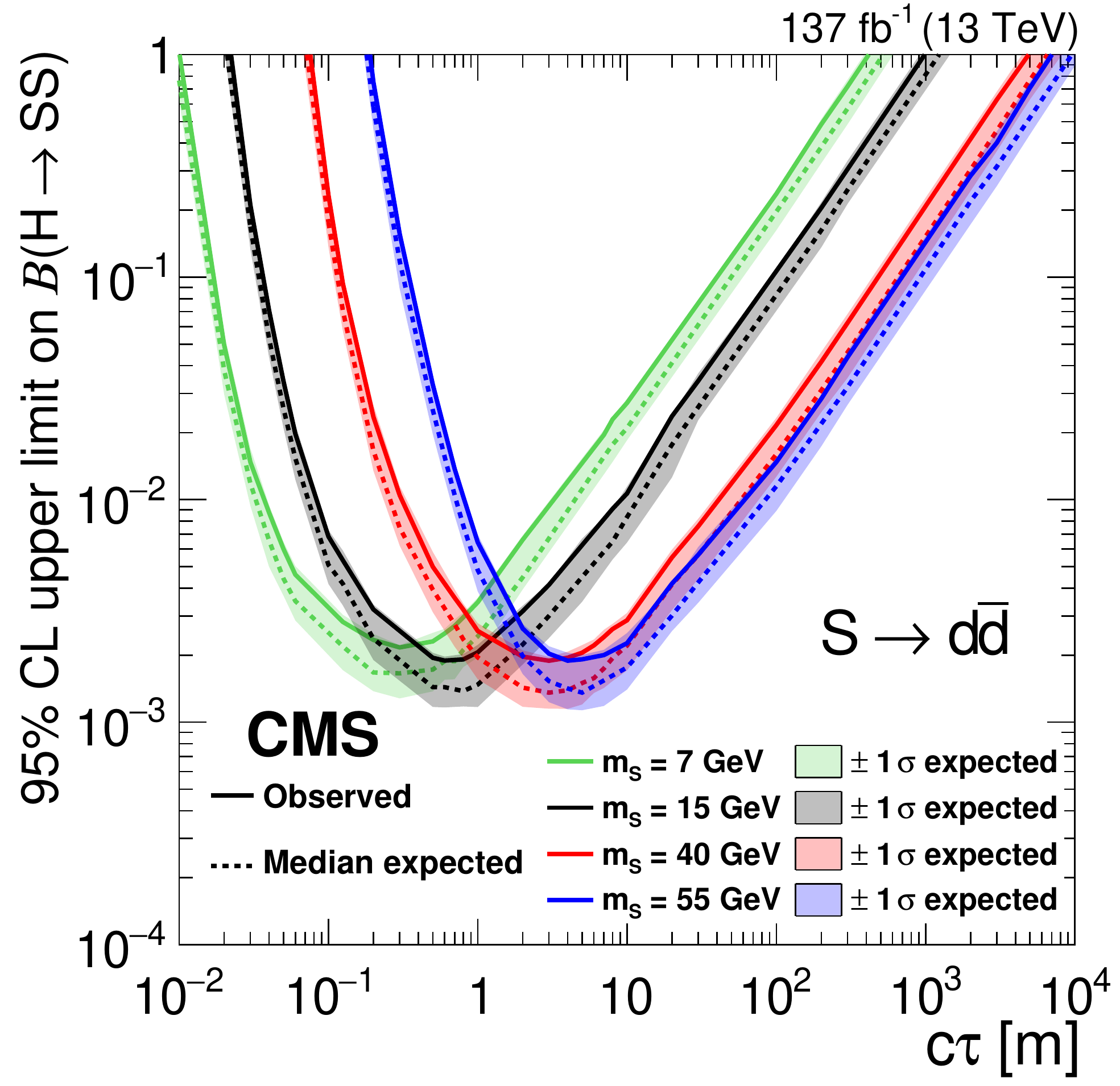}
  \includegraphics[width=0.49\textwidth]{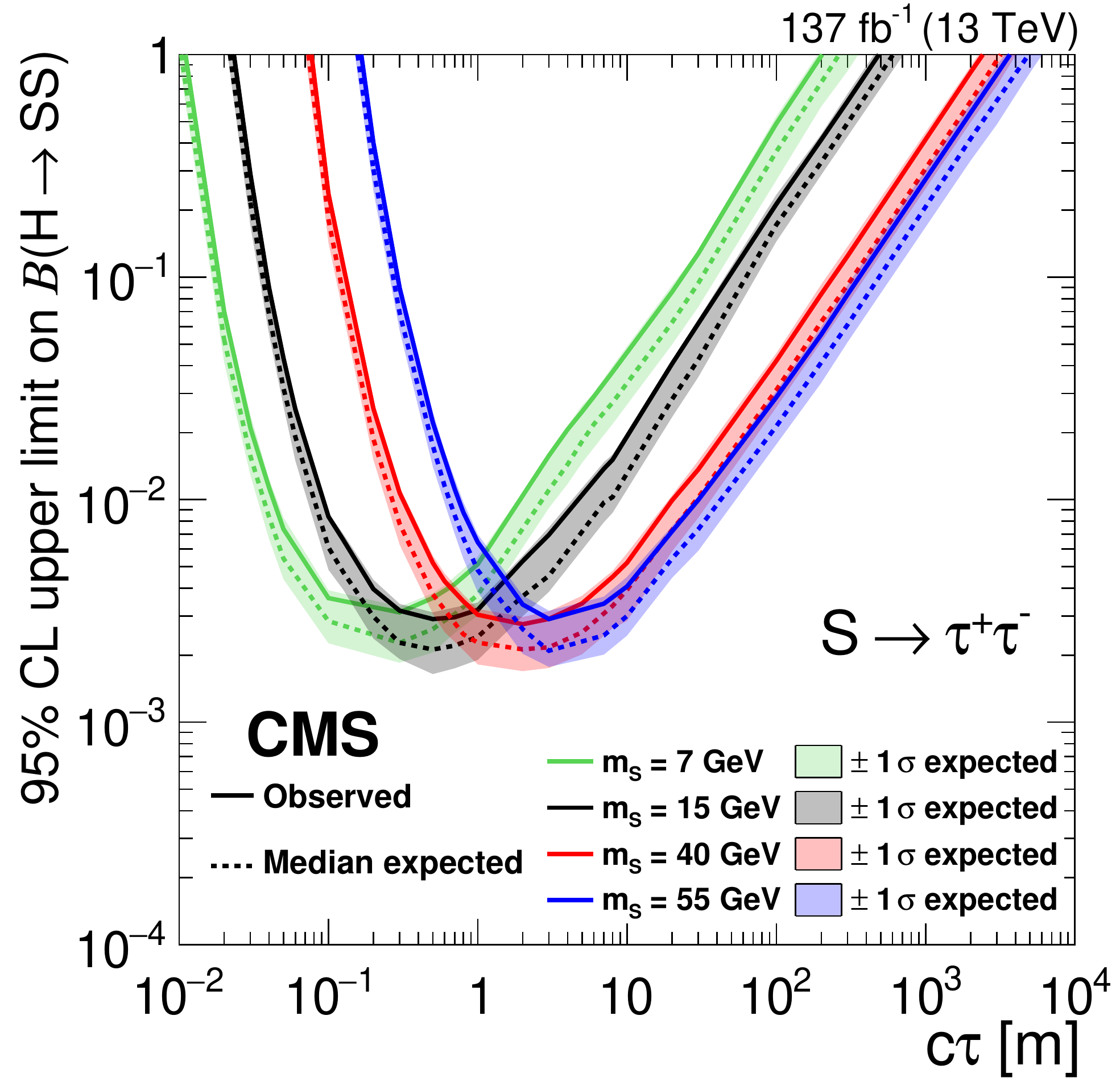}

  \caption{\label{fig:limit_vbfh}
  The 95\% \CL expected (dotted curves) and observed (solid curves) upper limits on the
branching fraction $\brhtoss$ as functions of $c\tau$ for the $\PS\to\ddbar$ (\cmsLeft) and
$\PS\to\PGtp\PGtm$ (\cmsRight) decay modes. The exclusion limits are shown for
four different mass hypotheses: 7, 15, 40, and 55\GeV. }
\end{figure}

We evaluate 95\% confidence level (\CL) limits on the branching fraction $\brhtoss$ using
the modified frequentist criterion \CLs~\cite{JUNK1999435,Read_2002,CMS-NOTE-2011-005} with
the profile likelihood ratio test statistic. The upper limits are shown in
Fig.~\ref{fig:limit_vbfh} for the $\PS\to\ddbar$ and
$\PS\to\PGtp\PGtm$ decay modes, as a function of $c\tau$ for a selection of
values of $m_\PS$. The exclusion limits for
$\PS\to\bbbar$ are within 3\% of
the exclusion limits for $\PS\to\ddbar$ for $m_\PS > 2m_{\Pb}$.

In summary, proton-proton collision data at $\sqrt{s} = 13\TeV$ recorded by the
CMS experiment in 2016--2018, corresponding to an
integrated luminosity of 137\fbinv, have been used to conduct the first
search for beyond the standard model (SM) long-lived particles (LLPs)
using the CMS endcap muon detectors as a calorimeter.
Based on a unique detector signature, the search is largely model-independent, with sensitivity to a broad range of LLP decay modes and to LLP masses as low as a few GeV.
With the excellent shielding provided by the inner CMS
detector, the background is suppressed to a low level and a search
for a single LLP decay is possible.
No significant deviation from the SM background is observed, and the most stringent
limits on the branching fraction of Higgs boson to LLP decaying to $\ddbar$, $\bbbar$, and $\PGtp\PGtm$ are set for proper decay lengths $c\tau > $
6, 20, and 40$\unit{m}$, and LLP masses of 7, 15, and 40\GeV, respectively.
For $c\tau > 100\unit{m}$, this search outperforms the
previous best limits~\cite{Aad:2019xav,Aaboud:2018aqj} by a factor of
6\,(2) for an LLP mass of 7\,($\ge$15)\GeV.

\begin{acknowledgments}
   We congratulate our colleagues in the CERN accelerator departments for the excellent performance of the LHC and thank the technical and administrative staffs at CERN and at other CMS institutes for their contributions to the success of the CMS effort. In addition, we gratefully acknowledge the computing centers and personnel of the Worldwide LHC Computing Grid and other centers for delivering so effectively the computing infrastructure essential to our analyses. Finally, we acknowledge the enduring support for the construction and operation of the LHC, the CMS detector, and the supporting computing infrastructure provided by the following funding agencies: BMBWF and FWF (Austria); FNRS and FWO (Belgium); CNPq, CAPES, FAPERJ, FAPERGS, and FAPESP (Brazil); MES (Bulgaria); CERN; CAS, MoST, and NSFC (China); MINCIENCIAS (Colombia); MSES and CSF (Croatia); RIF (Cyprus); SENESCYT (Ecuador); MoER, ERC PUT and ERDF (Estonia); Academy of Finland, MEC, and HIP (Finland); CEA and CNRS/IN2P3 (France); BMBF, DFG, and HGF (Germany); GSRT (Greece); NKFIA (Hungary); DAE and DST (India); IPM (Iran); SFI (Ireland); INFN (Italy); MSIP and NRF (Republic of Korea); MES (Latvia); LAS (Lithuania); MOE and UM (Malaysia); BUAP, CINVESTAV, CONACYT, LNS, SEP, and UASLP-FAI (Mexico); MOS (Montenegro); MBIE (New Zealand); PAEC (Pakistan); MSHE and NSC (Poland); FCT (Portugal); JINR (Dubna); MON, RosAtom, RAS, RFBR, and NRC KI (Russia); MESTD (Serbia); SEIDI, CPAN, PCTI, and FEDER (Spain); MOSTR (Sri Lanka); Swiss Funding Agencies (Switzerland); MST (Taipei); ThEPCenter, IPST, STAR, and NSTDA (Thailand); TUBITAK and TAEK (Turkey); NASU (Ukraine); STFC (United Kingdom); DOE and NSF (USA).
\end{acknowledgments}

\bibliography{auto_generated}
\cleardoublepage \appendix\section{The CMS Collaboration \label{app:collab}}\begin{sloppypar}\hyphenpenalty=5000\widowpenalty=500\clubpenalty=5000\vskip\cmsinstskip
\textbf{Yerevan Physics Institute, Yerevan, Armenia}\\*[0pt]
A.~Tumasyan
\vskip\cmsinstskip
\textbf{Institut f\"{u}r Hochenergiephysik, Wien, Austria}\\*[0pt]
W.~Adam, J.W.~Andrejkovic, T.~Bergauer, S.~Chatterjee, M.~Dragicevic, A.~Escalante~Del~Valle, R.~Fr\"{u}hwirth\cmsAuthorMark{1}, M.~Jeitler\cmsAuthorMark{1}, N.~Krammer, L.~Lechner, D.~Liko, I.~Mikulec, P.~Paulitsch, F.M.~Pitters, J.~Schieck\cmsAuthorMark{1}, R.~Sch\"{o}fbeck, D.~Schwarz, S.~Templ, W.~Waltenberger, C.-E.~Wulz\cmsAuthorMark{1}
% ...
\vskip\cmsinstskip
% ...
1:  Also at TU Wien, Wien, Austria\\
% ...
\end{sloppypar}
\end{document}